# Hot isostatic pressing of bulk magnesium diboride: mechanical and superconducting properties

S.S. Indrakanti,[a] V.F. Nesterenko,[a,d*] M.B. Maple,[b,c] N.A. Frederick,[b,c]

W.M. Yuhasz,[c,d] Shi Li[b,c]

[a]Department of Mechanical and Aerospace Engineering; [b]Department of Physics; [c]Institute for Pure and Applied Physical Sciences; [d]Materials Science and Engineering Program; University of California, San Diego, La Jolla, CA 92093, USA

**ABSTRACT**

Two different hot isostatic pressing cycles (HIPing) were investigated to synthesize bulk $MgB_2$ samples: a "standard" cycle where a low vessel pressure is maintained while heating to the process temperature with a subsequent simultaneous pressure and temperature decrease and a new method - dense material cooling under pressure (DMCUP). The latter method allowed the synthesis of dense samples with diameters up to 20 mm and thicknesses up to 10 mm from commercial $MgB_2$ powder. Optimal conditions for the DMCUP method with glass encapsulation (maximum pressure 200 MPa, maximum temperature 1000 °C over 200 min, and cooling under pressure) resulted in a dense material with a sharp superconducting transition at 38.5 K. This method employs a pressure which is one order of magnitude less than previously reported for pressure assisted sintering of dense material and can be scaled to larger sample sizes and complex shapes. The data for density, microhardness, fracture toughness and sound speed as well as superconducting properties for bulk magnesium diboride are presented. Ball milling the powder enhances sintering and results in a more homogeneous final microstructure.

## 1. Introduction

A relatively high temperature superconducting transition was recently found in the non-oxide compound $MgB_2$ [1,2]. Due to the particular properties of this material such as brittleness and decomposition at relatively low temperatures, about 1000 °C, fully dense solid samples required for many applications and for fundamental investigations of the superconducting properties are difficult to synthesize.

Superconductivity in $MgB_2$ was discovered in material prepared by Nagamatsu et al. from a mixture of Mg and B powder [1]. Pressed pellets were hot isostatically pressed at 700 °C in an argon atmosphere at a pressure of 196 MPa for 10 hours. No encapsulation step was reported. The behavior of the magnetic susceptibility of the sample below the superconducting transition is similar to that of the commercially available $MgB_2$ powder [2,3,4]. Using $MgB_2$ powder (Alfa Aesar, Inc) as a starting material, Jung et al. [2,3] and Takano et al. [4] employed a 12-mm cubic multi-anvil-type press at a pressure of 3000 MPa and a temperature of 950 °C to synthesize samples with diameters of ~4.5 mm and heights of ~3.3 mm. It is not clear why such a high level of pressure was required to synthesize a fully dense sample and the natural limitations of a multi anvil press preclude the use of this method for applications requiring samples of large size and complex shape. Therefore it is important to develop a method which will allow scaling of the sample sizes and synthesis of complex shapes.

The main goal of our research is the experimental investigation of the possibility of forming dense samples of $MgB_2$ with high quality superconducting properties under high pressure, which will also make the synthesis of large scale samples of complex shape possible. An additional objective is to investigate the effect of high energy ball milling (SPEX 8000) of the initial powder on the microstructure of the synthesized material and the influence it has on mechanical and superconducting properties.

It was demonstrated for other metallic and ceramic materials that high energy plastic deformation (mechanical milling, dynamic shock densification) combined with traditional sintering or HIPing can increase sinterability of powder, reduce final grain size, and improve mechanical properties. One of the reasons for this behavior is the increase of the driving force for the nucleation and the growth of new grains due to the defect structure created during high energy plastic deformation. Additional reasons can be connected with the increased diffusion transport caused by the mechanically induced defect structure and reduction of particle size, and with the enhanced contribution of grain boundary sliding during viscoplastic flow [5,6,7,8,9]. No results are published on the structural modification and superconducting properties of $MgB_2$ caused by high energy treatment in combination with sintering or HIPing.

## 2. Experiment

Magnesium diboride powders of –325 mesh size and 98% purity were obtained from Alfa Aesar, Inc. One batch of powder was milled in a SPEX 8000 ball mill with WC balls and a WC lined vessel for approximately 30 minutes with a balls to powder ratio of approximately 7:1 (mass of powder 5 g). Both milled and non-milled $MgB_2$ powders were separately placed in their own closed cylinders made from tantalum foil. Each cylinder was covered with a zirconium foil which served as a getter. The whole assembly was sealed in a pyrex capsule under a vacuum of $10^{-2}$ torr.

HIPing was carried out in an ABB Mini-HIPer using two different cycles: a "standard" HIPing cycle II [9] and the DMCUP method [10,11]. In the first method, the maximum temperature was 1000 °C and the pressure was 200 MPa with a subsequent simultaneous decrease of pressure and temperature (Fig. 1). The "standard" HIPing of mechanically milled powders in these conditions resulted in solid material with a gold color, but a macrocrack appeared in the middle part of the sample despite its relatively small size of about 5 mm diameter (Figure 2). HIPing of the non-milled commercial powder in the same cycle did not result in a solid compact although the material was densified (Fig. 3). We speculate that this dramatic difference is due to the enhanced diffusion caused by the defect structure of the ball milled powder and the reduction of the particle size. Further research is under way to explain this activation of sintering due to the high energy treatment of the $MgB_2$ powder.

The DMCUP HIPing procedure was applied under two conditions. In the first schedule the maximum temperature was 850 °C and the pressure was 200 MPa. The resulting samples were dense (density 2.39 g/cm$^3$ for milled and 2.49 g/cm$^3$ for non-milled powder, experimental error 3%), but contained local pores, and the superconducting transition based on magnetic susceptibility measurements was similar to that obtained for the initial powder.

The second condition of the DMCUP procedure, which resulted in the best mechanical and superconducting properties, used an increased temperature of 1000 °C with the same pressure of 200 MPa. A corresponding HIPing cycle for this schedule is presented in Figure 4. The microstructures of the sintered samples are presented in Figure 5. We can clearly see that ball milling of the powder results in a more homogeneous microstructure similar to that produced by the "standard" HIPing cycle but without macrocracking in the central part of the sample. Small cracks were present in the periphery of the sample and were caused by sharp corners on the Ta container wall.

The critical feature of this method which employs dense material cooling under pressure (DMCUP) [10,11] is the delay of the pressure decrease with respect to temperature (Figure 4). The temperature history in standard HIPing cycle and in DMCUP method is very similar (Figs. 1 and 4). We interpret the completely different results produced by the "standard" HIPing cycle and the DMCUP method for the "as is" and milled powders as the influence of the pressure in the latter method during the cooling stage which helps to prevent cracking under thermal gradients and effectively prolong pressure assisted sintering on the stage of cooling.

Despite the clear advantage of the DMCUP method in comparison to the "standard" HIPing cycle with a similar temperature cycle we do not consider the selected parameters for both approaches to be fully optimized. It is possible that a slow and coordinated decrease of temperature and pressure in the "standard" HIPing cycle could result in crack free and fully dense solid samples. Based on our experiments only a slight increase in the temperature hold of "as is" material can result in its sintering in the "standard" HIP cycle.

## 3. Results and discussion

### 3.1 Mechanical properties

Because the "standard" HIP cycle II at investigated conditions did not result in crack free solid samples we will only characterize materials obtained with the DMCUP method. The density of the synthesized material using the optimized DMCUP method was 2.666 g/cm$^3$ (experimental error $\pm 0.004$ g/cm$^3$) and appeared to be higher than the reported theoretical density based on X-ray measurements (2.625 g/cm$^3$) [12] and similar to the density reported in [4] after pessure assisted sintering at 3000 MPa. Measurements of density were made based on an ASTM B 328 standard by weighing samples immersed in water. The experimental error was found based on the comparison of the density measurements of pure aluminum samples (wire, 99.999% Al) with the known density of solid aluminum, 2.6989 g/cm$^3$ [13]. This comparison was made with the mass of the Al sample close to the mass of the MgB$_2$ sample; these two materials have very similar densities.

The DMCUP method allows us to sinter relatively large dense samples suitable for the sound velocity measurements. A sample with thickness 2.8 mm and diameter 16 mm and parallel sides was machined for this purpose. The sound speed was measured with a Panametrics pulser-reciever Model 5072 PR using a transducer of 0.125" diameter and 20 MHz frequency. The signals were detected using an oscilloscope (Tectronix Inc.). The instrument was calibrated using a standard 1018 steel test block supplied. The longitudinal sound speed for synthesized $MgB_2$ was found to be 6340 m/s.

The high density and strength of the synthesized material makes it possible to prepare a polished shiny surface of mirror quality suitable for microhardness measurements. It is important that the indentations are acceptable for the measurements of Vickers microhardness in brittle materials [14]. The average Vickers microhardness of the samples measured under a load of 1 kgf was equal to 10.4 GPa (diagonal length 2a = 41.8 µm) for the sample prepared from non-milled powder and 11.7 GPa (diagonal length 2a = 39.4 µm) for the sample from milled powder.

As shown in Fig. 6, during the microhardness measurements transgranular macrocracks (with average lengths of about 96.6 µm and 86.3 µm calculated from the tip to tip (marked as 2c) for non-milled and milled samples respectively) emanated from all four corners of the indentation, which is a typical feature for brittle materials. The presented indentation's shape is typical for fully dense ceramic materials. These types of cracks satisfying the condition c ≥ 2a can be used to estimate the fracture toughness using the value of Young's modulus and based on the semiemperical equation [15] $K_C = 0.0226\,(EP)^{1/2}\,a\,c^{-3/2}$, where $E$ is Young's modulus, $P$ is the load used to produce the indentation (9.807 N), and $a$ and $c$ are the lengths of the diagonal and the crack as shown in Fig. 6. The Young's modulus of the sample (79.6 GPa) was estimated based on measured longitudinal sound speed and density, taking Poisson's ratio as a first approximation to be equal to 0.3.

The calculated fracture toughness $K_C$ was found to be 1.26 MPa·m$^{1/2}$ and 1.40 MPa·m$^{1/2}$ for samples HIPed from non-milled and milled powders, respectively. This value of fracture toughness is between the value for monocrystalline silicon and monocrystalline Al$_2$O$_3$ (sapphire) [15].

**3.2 Superconducting properties**

DC magnetic susceptibility $\chi_{dc}$ measurements were made with a commercial Quantum Design SQUID magnetometer in a magnetic field of 10 Oe. First, the sample was cooled in zero magnetic field from above its superconducting transition temperature down to 5 K, and then a magnetic field of 10 Oe was applied. Zero field cooled (ZFC) susceptibility data were thus obtained by measuring the magnetization as the sample was warmed to 60 K. Field cooled (FC) data were collected from magnetization measurements as the sample was cooled down to 5 K with the magnetic field still applied. Resistivity measurements were made using a standard four-probe technique in a commercial Quantum Design Physical Properties Measurement System (PPMS).

The superconducting characteristics of untreated powder and three HIPed samples are compared in Figure 7 as plots of $\chi_{dc}$ vs. T where the data were normalized to their values at 5 K and 60 K. In all of the samples there is a considerable difference between the FC and the ZFC curves, consistent with the pinning expected in a Type II superconductor. However, the initial powder (sample A) and the sample HIPed at 850 °C (sample B) (see Figure 7) also show a significant magnetic moment in the FC data. In contrast, the two samples HIPed at 1000 °C, samples C and D (non-milled and milled, respectively; see Figure 7), exhibit no diamagnetism in their FC curves. The crucial impact of the full density and cohesivness of the samples prepared by the DMCUP method at 1000 °C on the pinning strength can be clearly distinguished in comparison with the sample from the initial powder and the sample HIPed at 850 °C.

Electrical resistivity ρ vs. T data for both of the samples HIPed at 1000 °C (samples C and D) are shown in Figure 8 between 7 K and 300 K. The shapes of the ρ(T) curves for these samples above the superconducting transition are reminiscent of a typical metal. The sample HIPed at 850 °C (sample B) exhibited a prohibitively large resistance at room temperature, and thus was not measured. A residual resistivity ratio (RRR) between 300 and 40 K for samples from the as is powder was found to be 3.6 which is similar to the RRR = 3 for samples synthesized from commercial poweder at 3000 MPa [3]. For samples prepared from milled powders this ratio was equal to 2.7 with a higher resistivity at 300 K. This difference may be interpreted as caused by smaller grain sizes in the sample HIPed from milled powder.

The superconducting transitions of the two samples HIPed at 1000 °C and 200 MPa using the DMCUP method are extremely sharp in both dc magnetic susceptibility and electrical resistivity data (Figures 7 and 8). These measurements are consistent with those obtained on samples prepared at pressures of 30 kbar in a multi-anvil-type press [2,3,4] and improved over those attained for the HIPed samples of $MgB_2$ (probably without encapsulation) synthesized by Nagamatsu et al. [1]. The superconducting properties of the materials synthesized in this work under various treatments are shown in Table 1. We define the transition temperature $T_C$ as the temperature at which the susceptibility or resistivity is half of its final value and the transition width $\Delta T_C$ as the difference between the temperatures at 10% and 90% of the final value. Further investigation needs to be done to determine whether the difference in $T_C$ and $\Delta T_C$ between the samples HIPed at 1000 °C is due to the ball milling or is simply within experimental error.

Additional data on the superconducting properties of samples prepared from non-milled powder are presented in a separate paper [16].

4. Conclusions

In summary, dense, bulk, crack free materials exhibiting sharp superconducting transitions were synthesized using the dense material cooling under pressure (DMCUP) method based on hot isostatic pressing, which can be utilized at significantly lower pressures (200 MPa) than previously reported. The superconducting properties are similar to those of samples synthesized at 3000 MPa using a multi anvil press. In comparison with the latter method, our method allows scaling of the size of the samples and the manufacture of complex shapes. The mechanical properties (density 2.666 g/cc, sound speed 6340 m/s, microhardness 10.4 and 11.7 GPa, fracture toughness 1.26 and 1.40 MPa·m$^{1/2}$ for samples sintered from non-milled and milled powders, respectively) were experimentally measured.

**Acknowledgements**

This research was supported by the U.S. Department of Energy under Grant No. DE-FG03-86ER-45230.

Table 1

Superconducting transition characteristics determined from dc magnetic susceptibility $\chi_{dc}$ and electrical resistivity $\rho$ measurements for Alfa Aesar $MgB_2$ powder under various treatments. $T_C$ is defined as the temperature at which the susceptibility or resistivity is 50% of its final value. $\Delta T_C$, the width of the superconducting transition, is defined as the difference between the temperatures at 10% and 90% of the final value.

| **Alfa Aesar $MgB_2$ powder** | | $\chi_{dc}$ | | $\rho$ | |
|---|---|---|---|---|---|
| Sample | Treatment | $T_C$ (K) | $\Delta T_C$ (K) | $T_C$ (K) | $\Delta T_C$ (K) |
| A | As received | 36.79 | 12.54 | — | — |
| B | HIPed @ 850 °C | 36.75 | 10.57 | — | — |
| C | Non-milled, HIPed @ 1000 °C | 37.66 | 0.75 | 38.50 | 0.85 |
| D | Milled, HIPed @ 1000 °C | 37.24 | 0.35 | 38.11 | 0.76 |

**Figure captions**

Figure 1. Temperature and pressure histories of the "standard" hot isostatic pressing cycle II.

Figure 2. Cracking of samples sintered from ball milled powder in the conditions of the "standard" HIPing cycle II.

Figure 3. Nonsintered sample from commercial powder after the "standard" HIPing cycle II, under the same conditions as for the HIPing of the ball milled powder.

Figure 4. Temperature and pressure histories of the hot isostatic pressing cycle for the DMCUP method.

Figure 5. Microstructures of the "as is" (a) and milled (b) powders after DMCUP method.

Figure 6. Indentations after measurements of Vickers microhardness of "as is" (a) and milled powders (b). Note cracks emanating from the indentation corners.

Figure 7. DC magnetic susceptibility $\chi_{dc}$ (normalized to values at 5 K and 60 K) versus temperature of Alfa Aesar $MgB_2$ samples. Shown are measurements on the original powder and on samples HIPed using the DMCUP method.

Figure 8. Electrical resistivity ρ vs. T of Alfa Aesar $MgB_2$ HIPed at 1000 °C using the DMCUP method.

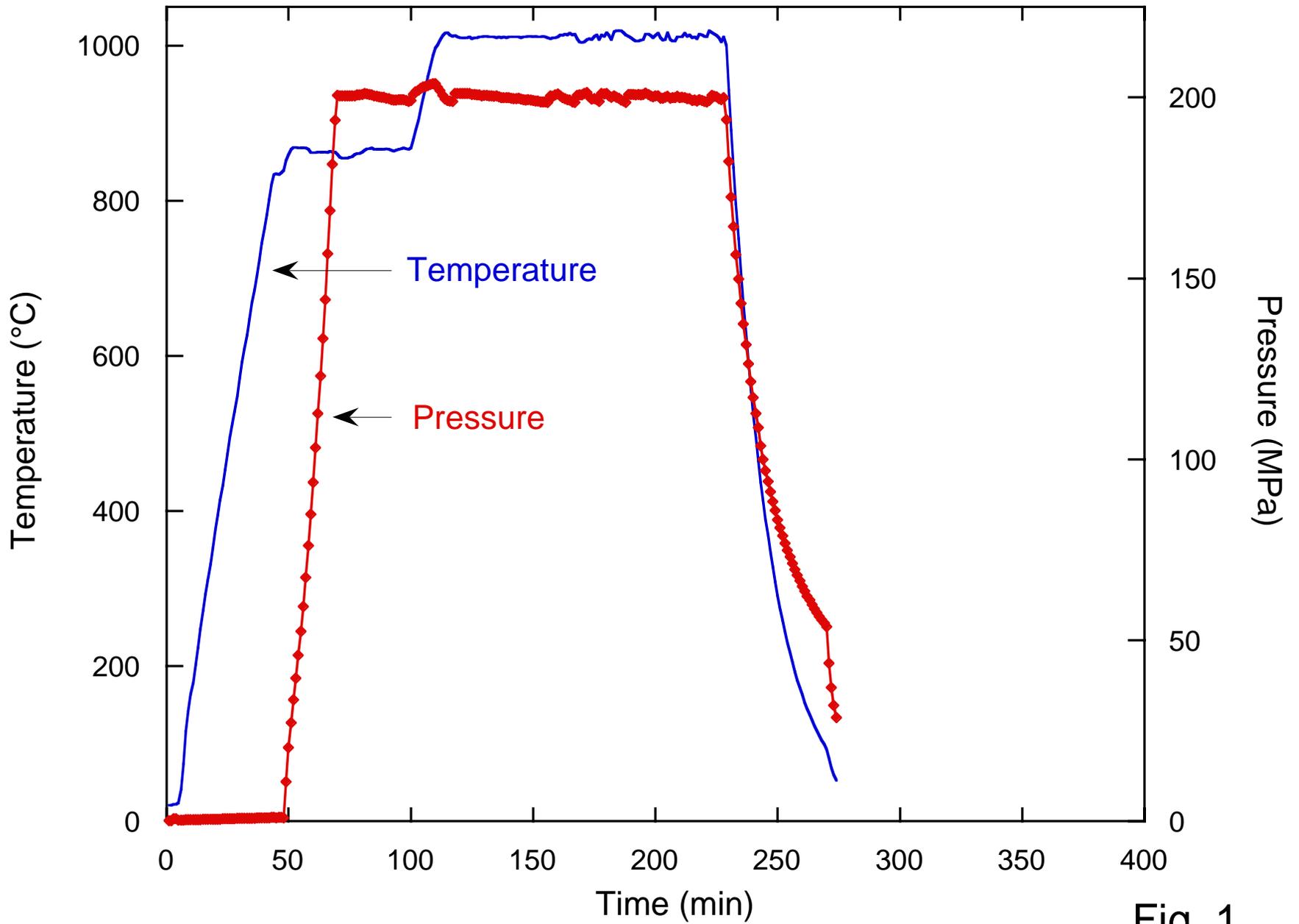

Fig. 1

# Fig. 2

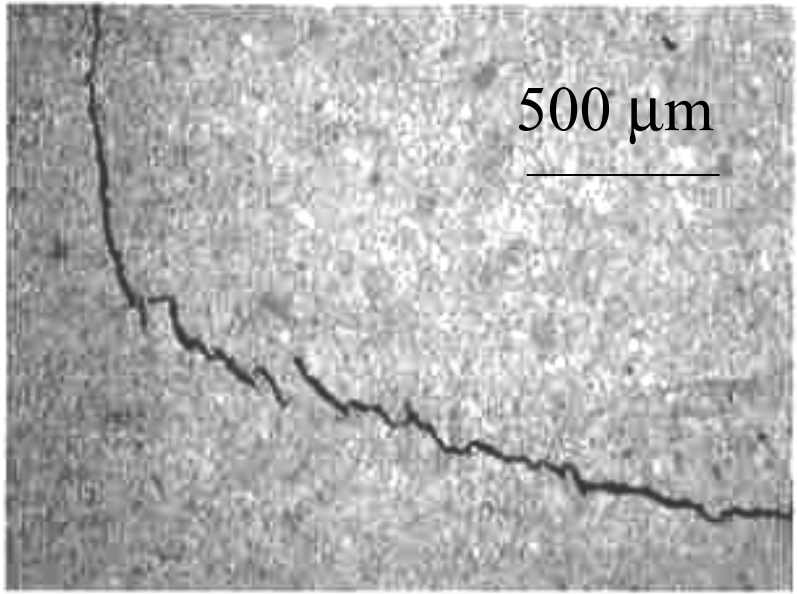

# Fig. 3

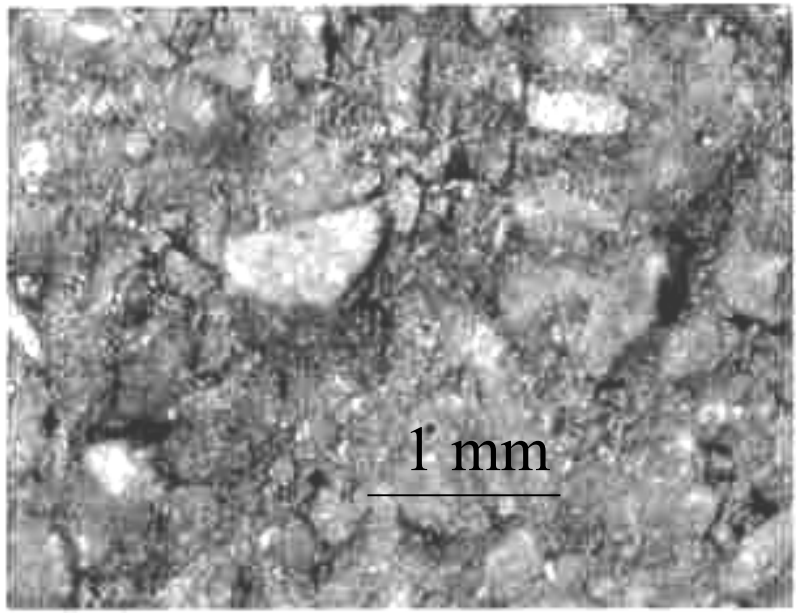

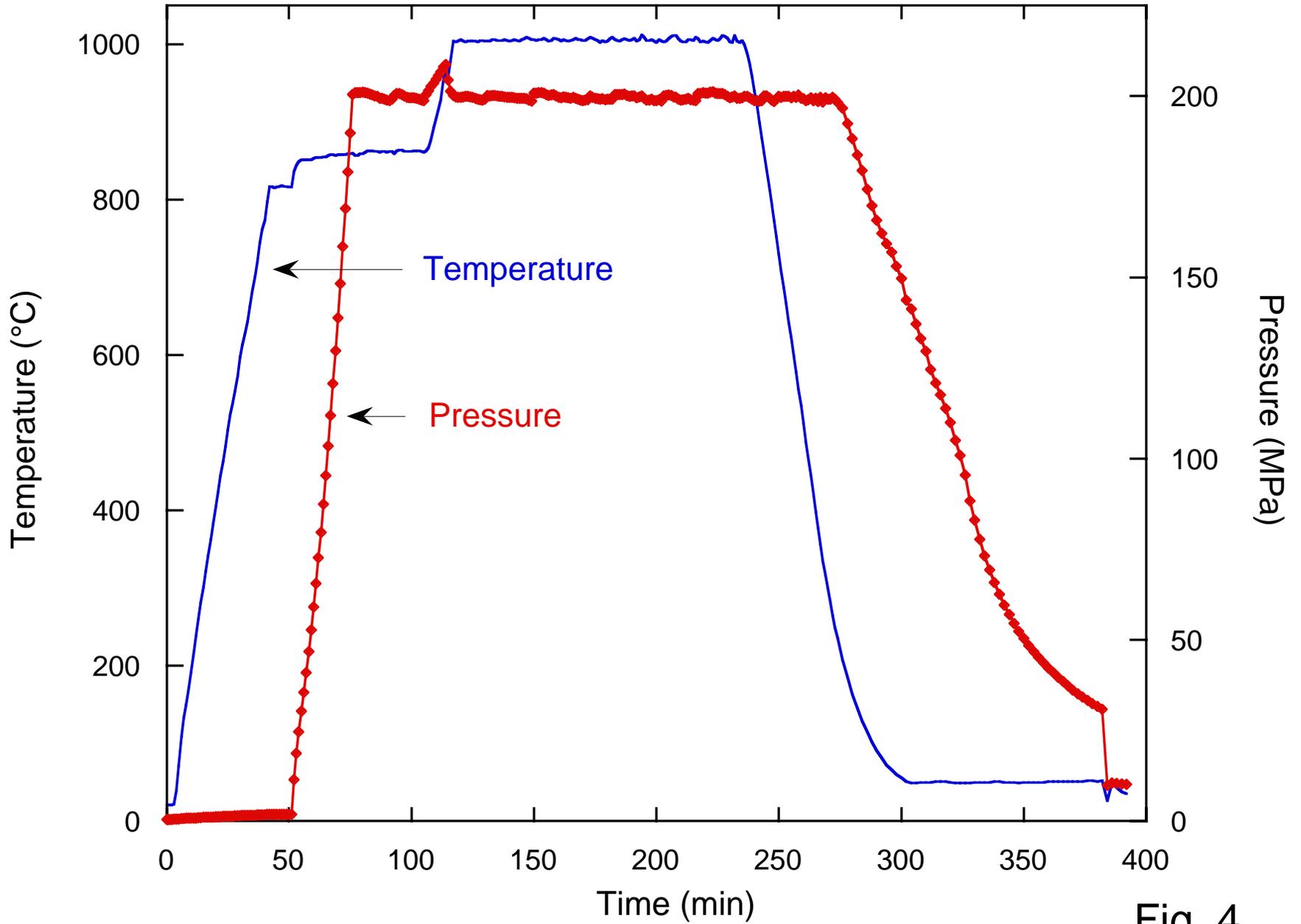

Fig. 4

Fig. 5a

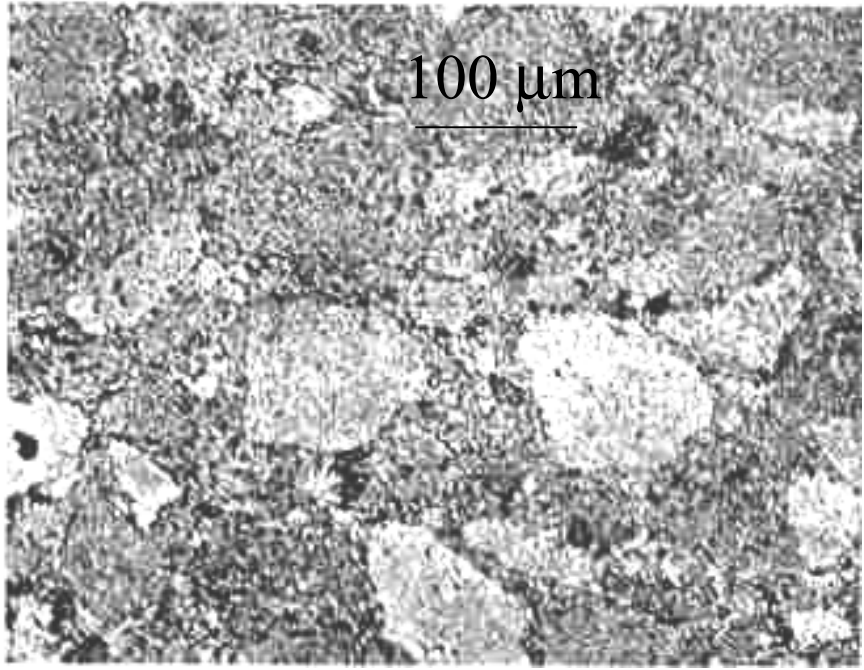

Fig. 5b

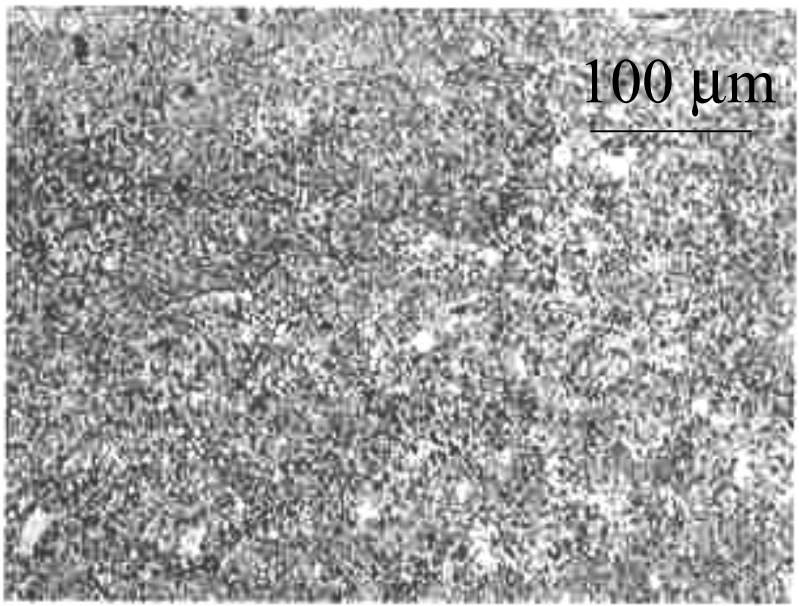

Fig. 6a

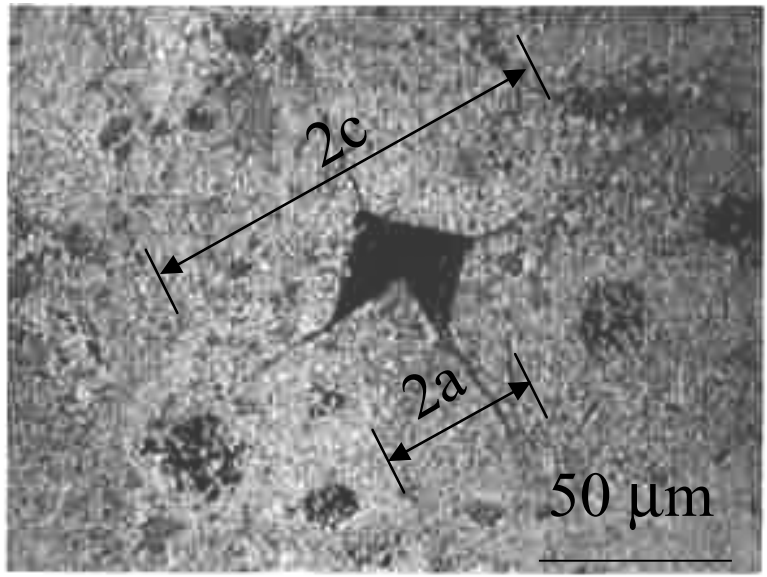

Fig. 6b

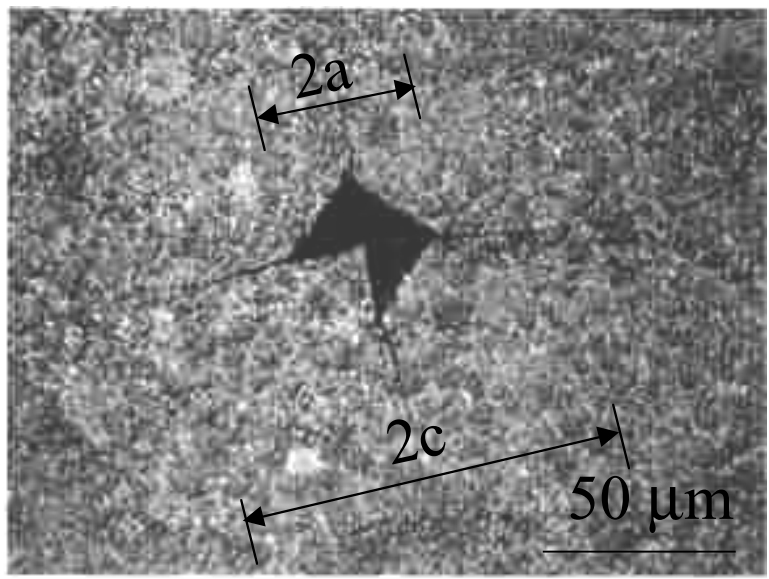

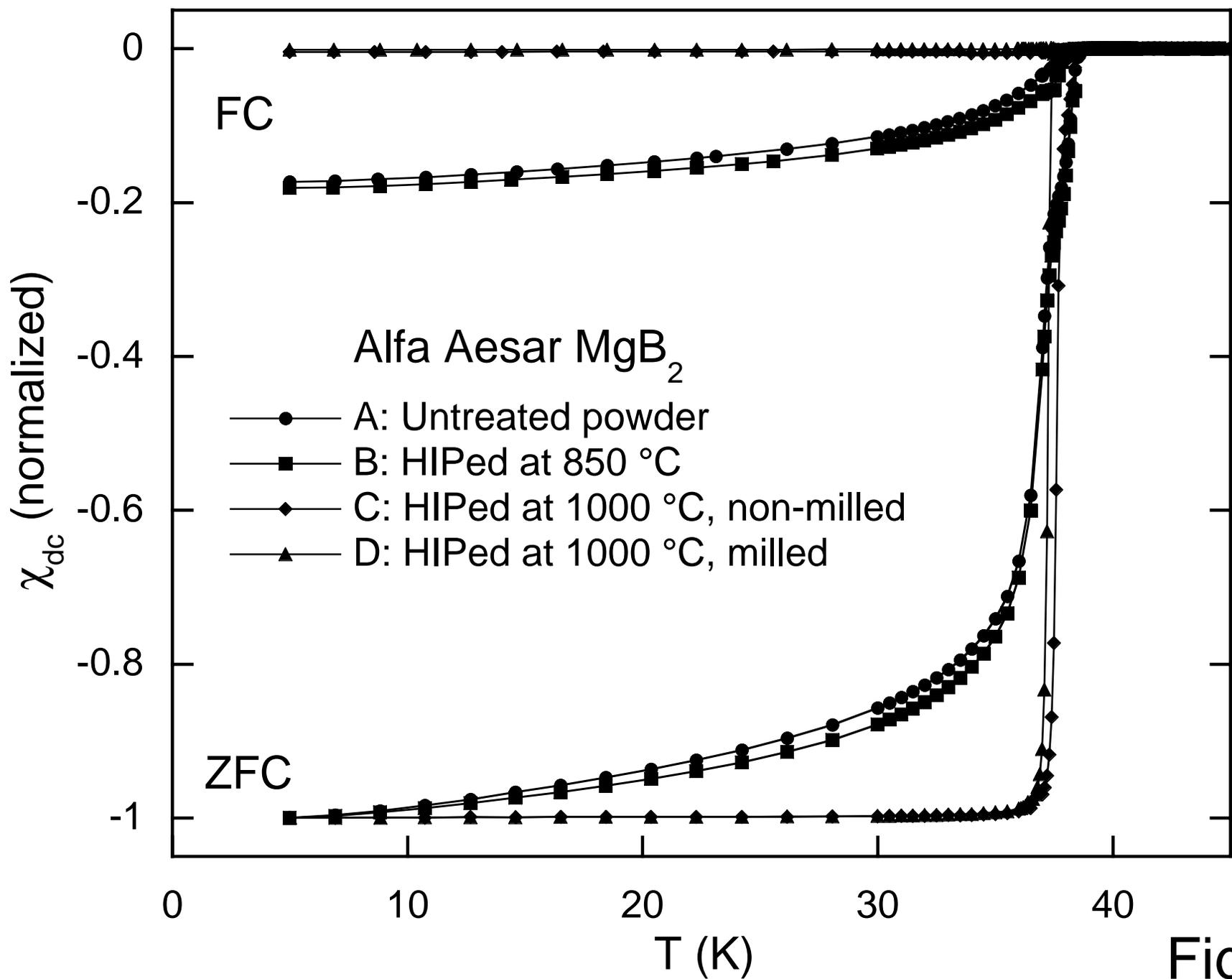

Fig.7

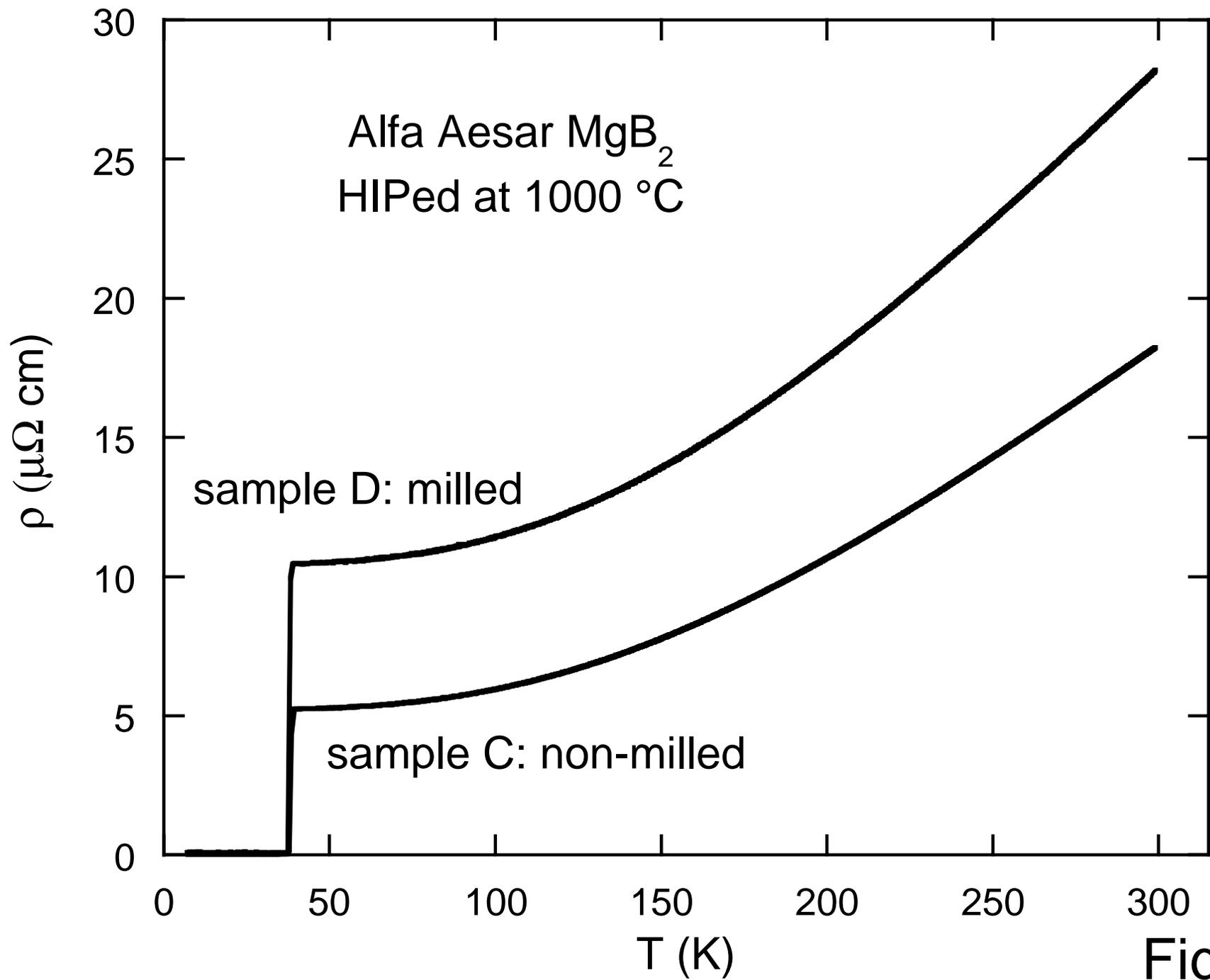

Fig.8